# Toward Evolution Strategies Application in Automatic Polyphonic Music Transcription using Electronic Synthesis

Herve Kabamba Mbikayi
Département d'Informatique de Gestion
Institut Supérieur de Commerce
Kinshasa, The Democratic Republic of Congo

*Abstract*—We present in this paper a new approach for polyphonic music transcription using evolution strategies (ES). Automatic music transcription is a complex process that still remains an open challenge. Using an audio signal to be transcribed as target for our ES, information needed to generate a MIDI file can be extracted from this latter one. Many techniques presented in the literature at present exist and a few of them have applied evolutionary algorithms to address this problem in the context of considering it as a search space problem. However, ES have never been applied until now. The experiments showed that by using these machines learning tools, some shortcomings presented by other evolutionary algorithms based approaches for transcription can be solved. They include the computation cost and the time for convergence. As evolution strategies use self-adapting parameters, we show in this paper that by correctly tuning the value of its strategy parameter that controls the standard deviation, a fast convergence can be triggered toward the optima, which from the results performs the transcription of the music with good accuracy and in a short time. In the same context, the computation task is tackled using parallelization techniques thus reducing the computation time and the transcription time in the overall.

## I. INTRODUCTION

Automatic music transcription is the process that involves a computer in order to write partitures of a piece of music or an audio signal. In automatic music transcription, a piece of music or an audio signal is analysed in order to figure out the correspondent human representations of the perceived sound for a proper interpretation. As humans, we sense sound by noticing minute differences in the pressure of the air outside of our ears, hence these intensities can be reproduced by speakers and moreover they can be recorded by microphones.

The audio signal can be composed of a succession of monophonic sounds over the time which involves a set of notes played individually but not at the same time; or polyphonic sounds which are a succession of notes played at the same time from the same instrument or even from different instruments.

A music transcriber program should be able to analyse the audio signal in order to figure out the different elements composing the given signal. Transcribing music is a very difficult problem that still remains unsolved in a computational view, however even in a musical view this knowledge belongs only to experimented and skilled musicians who have not only learned to transcribe melodies but who have developed their internal skills over the time through their experiences or their predispositions. A formal way to transcribe music is to represent this information as universal sheets of music containing specific information on what notes are played at a given time without saying what the result will sound like. Using a computer, a way to represent music information is to use for instance MIDI files which abstract most of the information composing a piece of music in order to interpret it on a computer or in the real world. Music transcription mainly relies on the use of digital signal processing techniques, but current approaches are not able to capture the rich diversity found in audio signal [15], and some approaches propose the use of semi-automatic transcription to get better results [16].

The transcription problem can be also seen as a search space problem which consists to find optimal or acceptable representations of the music signal which are likely to be used directly or even as a starting point to find the optimal information. Due to the huge size of the search space, the use of evolutionary algorithms (EAs) in this context is the correct approach to deal with it since these algorithms start running with a small set of information going through the search space toward the optimal values.

We introduce in this paper a new approach for music transcription using evolution strategies (ES), which are considered as belonging to the class of EAs. Although in the literature, some reported works used genetic algorithms (GA) to tackle this problem, this process being complex due the size of search space some questions are raised on the time needed for a full convergence to the optimum when GA are used. GA use binary representations to encode the individuals of a population. Operations on the different individuals that include recombination and mutation are applied on the chromosomes by means of permuting the positions of bits randomly or modifying the value of bit when mutation is applied, and exchanging set of bits information or blocks between two individuals when it comes to recombination.

The use of ES can be more adapted for such problem since, the optimization process is done by evaluating information which is presented in a way (frequencies, magnitudes, pitch, etc.) more adapted to deal with ES which contrarily to GA, use real-valued representations. Moreover, ES use self-adapting parameters which adapt themselves during the evolution process. The experiments showed that by tuning these parameters the time needed for convergence toward the optimal values can be significantly reduced while the obtained results still remain acceptable.

The rest of the paper is organized as follows: In section 2, an overview on ES is presented, in section 3 related work on the subject is discussed, in section 4 we discuss about the application of ES in music transcription, in section 5 we present and discuss about our experimental results while in section 6, we draw some conclusions talk about some future directions.

## II. EVOLUTION STRATEGIES

ES are a sub-class of nature-inspired direct search methods belonging to the class of EAs which use mutation, recombination, and selection applied to a population of individuals containing candidate solutions in order to evolve iteratively better and better solutions [1]. They can be applied in all fields of optimization including continuous, discrete, combinatorial search spaces without and with constraints as well as mixed search spaces. A first type of strategy includes directly the mutation strength for each attribute of an individual inside the individual. This mutation strength is subject to evolution similarly to the individual in a classic genetic algorithm. The encoded individual for this type of ES must include a strategy parameter as part of the individual. Rechenberg [9] and Schwefel [10], was the first to introduce them for optimizing real-vectors. They are considered as belonging to the class of EAs. Contrarily to GAs, ES do not use a binary representations to encode a chromosome. They use real-valued vectors to encode an individual. The emphasis is given to the mutation operator which applies to the chromosome a random noise from a normal distribution.

ES abstract the evolution process by which genes are affecting the phenotype of individuals and are the external expression of those genes within an individual. The presumption for coding the variables in the ES is the realization of a sufficient strong causality (small changes of the cause must create small changes of the effect) [2]. Parameters in an ES self-adapt as the evolution process is happening. They use a particular formalism to denote their different types. The simplest form of ES can be expressed as (1+1)-ES in which we have one parent $x^{(t)}$ that will produce one offspring $x'^{(t+1)}$ by mean of mutation. Offspring need sufficient level of adaptability to survive to the next generation. A fitness value should be assigned to them in order to evaluate their adaptability level with respect to the problem we are dealing with. The individual with the best fitness value is selected for the next generation t+1 as the selection operation is deterministic. Equation 1 from [2], describes this process.

$$x^{(t)} = \begin{cases} x'^{(t+1)} & if \ fi(x'^{(t+1)}) \geq fi(x^{(t)}) \\ x^{(t)} & otherwise \end{cases} \quad (1)$$

Based on the way the selection operation is done, ES are differentiated as (1+ λ)-ES called "plus selection ES" in which the selection of best individuals is applied on both the parents and offspring, and (1, λ)-ES called "comma selection ES" in which the selection happens only on the offspring. The formalisms (μ/ρ, λ)-ES, (1+ λ)-ES, (1, λ)-ES are also used to describe ES. In the latter one, the symbol μ represents the total number of parents, ρ represents the number of parents that will be recombined and, and λ stands for the number of offspring.

$$x'^{(t+1)} = x^{(t)} + \sigma^{(t)} . N(0,1) \quad (2)$$

The mutation operation is performed on $x^{(t)}$ using a random value $N(0, \sigma^{(t)})$ from a normal distribution with a mean of zero by applying the 1/5th-rule based on the rate of successful mutations. Successful mutations happen only when the fitness of the produced offspring is better than its parent while the global step-size ($\sigma^{(t)} \in \mathcal{R}+$) used during the mutation is itself adapted (Eq(2)).

$$\sigma^{(t+1)} = \begin{cases} \sigma^{(t)} . \alpha_{ES} & if \ fi(x'^{(t+1)}) \geq fi(x^{(t)}) \\ \sigma_{ES}^{-\frac{1}{4}} & otherwise \end{cases} \quad (3)$$

The normal distribution with a mean of zero and standard deviation of 1 is represented by N (0,1) whereas $\alpha_{ES}$ is the change rate of the global step-size. It is recommended to use values between $2^{1/n}$ and 2 for $\alpha_{ES}$ [2], where n is the dimension of the problem. Each element $x_i(t)$ should be initialized to a value $x_i(0)$ and α(t) to a constant value and α(0) the value for this constant depends on the problem. $\alpha_{ES}$ represents the changing rate of the step-size [2].

## III. RELATED WORK

In the literature many works addressing music transcription have been presented. They all use different approaches to address this issue. In [3], using a degenerative process, pitches was estimated by the subtraction of note estimate from the frequency spectrum at a time. Although this model presented good results, it suffers from Imperfections in terms of results obtained when it comes to dealing with small perturbations in the audio signal, as they result in the imperfect subtractions that cause problems with the whole transcription.

Masataka used an approach with every possible fundamental frequencies that could be guessed, and was able to transcribe a melody and the bass line among any numbers of instruments with good results [7]. In [8], Bayesian probability networks were used where bottom-up signal analysis could be integrated with temporal and musical predictions as polyphonic music can be addressed in a certain sense by considering the principles of human auditory organization. This probabilistic Bayesian networks was also used by Manuel Davy and Simon J. Godsill with variable-weight sound model as heuristic to transverse the large parameter space of music transcription [4]. This method however was only able to transcribe three simultaneous notes but could not do more.

The apparition of the first work applying evolutionary algorithms for music transcription was in 2001, in which Garcia [12], showed that polyphonic pitch detection can be seen as a search space problem in which the goal is to find the pitches that compose an acoustic signal. He used GAs for this purpose and instead of decomposing the frequency lattice which is almost impossible to do, his approach was to reconstruct it, thus considering it as a search space problem.

Jorge dos Reis and Francisco Fernandez de Vega were able to successfully use GAs to transcribe polyphonic music with electronic synthesis [6]. However this approach suffers from its computationally intensive behaviour and the convergence time which is very long due to the application of GAs in the context of polyphonic music transcription. In [5], GAs was also used for polyphonic music transcription and moreover, they were applied to address multi-timbral music transcription.

Gustavo Reis et al. also used GAs for automatic polyphonic music transcription by addressing the multiple fundamental frequencies problem and tracking [14]. Their approach was restricted to three cases which are: estimation of active fundamental frequencies on a frame-by-frame basis; tracking of note contours on a continuous time basis; tracking of timbre on a continuous time basis.

## IV. TECHNICAL APPROACH

Musical transcription is still an open challenge for researchers. Although there exists a lot of works that address this problem, most of them are trying to conceptualize this process which in reality is still unknown. There is no standard approach to transcribe music. Experimented musicians use rare internal skills that they possess and acquired over the time to tackle it whereas a formal way to develop them cannot be guaranteed as a lot of factors that are not likely to be totally abstracted, are involved in this process of developing internal abilities within a particular human. Standard methods are trying to quantify a process which is computationally unknown and even skilled musicians cannot provide formal methods to be used in order to transcribe music.

It is easy for a computer to go from a MIDI representation toward the synthesis of the specific audio signal; this representation can be understood by any program that can read a MIDI file. However going from an acoustic or an audio signal to a formal representation of that information raises concerns. This is a crucial issue that needs to be addressed in its context hence one has to discover who is playing what notes.

To tackle it, we developed a synthesizer that uses a particular MIDI-like representation to produce an audio signal during the transcription process. This synthesized audio signal is then compared with the original audio signal from a WAV file that needs to be transcribed. The ES takes as input this MIDI-like information containing standard midi numbers representing pitches to encode individuals of the population.

The evolution process then evaluates the level of adaptation of the population by comparing the synthesized signal with the original one from a Wav file.

In his work, Klapuri expressed the need for both a method for analysing the music and a mean of parameters for optimization [11]. A good optimization process applied to music transcription is likely to produce good results.

The fitness function comparing the two audio signals returns the results of the evaluation whereas the ES algorithm returns the improved hypotheses. With regards to the polyphonic music transcription, complex frequency lattice computationally infeasible to deconstruct is created. The benefit of this approach is that, this lattice will not need to be deconstructed but rather will be reconstructed. This reconstruction process will start from less fit individuals evolving toward fittest ones by means of ES operators applied to the population to get closer to ideal transcription of the music.

### A. Indivuduals Encoding

To encode individuals as primitives for our ES, a real-valued vector is used. This latter one contains a set of randomly generated real numbers that are part of the chromosome. The transcription uses a set of three notes to compose a sound at a given time, thus forming a chord. In this approach we are not taking into account the timbre of the instruments that needs to be transcribed although this can be easily done by improving our algorithm in order to analyse the timbre information of the signal and generate the correct envelopes with a synthesizer. However, we investigate on what notes can reconstruct the played signal as well as their durations. Table 1 shows a generated individual encoded as a chromosome.

TABLE I. ENCODED CHROMOSOME FOR AN INDIVIDUAL

| 59.98234 | 67.00242 | 64.03123 | 67.02345 | 71.36770 | 74.42356 |
|---|---|---|---|---|---|

The above example can be considered as a vector of 6 elements containing real values as they appear in the table. To encode a sound, we need three notes that should be played at the same time. Therefore, the example depicts a sequence of 2 sounds (each composed of three notes) played at different times. MIDI numbers are integers that represent pitches, the frequency of each pitches can be obtained using the formula given in equation 4. However a question is raised on the way ES would be applied in the context of evolving those chromosomes containing MIDI numbers which are integers.

The experiments have shown that by rounding the real number towards its closer midi integer gives very good and expected results. As per the example presented in Table 1, "59.9834" will be rounded to "60" and "67.00242" to "67". This rounding process is done only after the ES completed the optimization process and sends back the results. However depending on the context, one may need the algorithm to return real values as results for purposes of analysis or for a manual rounding which in some cases may improve the precision, and in some cases it is important to use directly these real numbers without any rounding process for transcription of music from acoustic instruments.

The first 3 numbers are respectively representing notes "C (60)", "G (67)", "E (64)". Played at the same time, they form a C major chord. The next three numbers respectively represent "G (67)", "B (71)", and "D (74)" and form a G Major when played at the same time.

$$Frequency = 6.875 \times 2^{\frac{3+\text{pitch}}{12}} \qquad (4)$$

Therefore, the optimization process will consist in finding the correct matching of notes to construct a signal that is closer to the target one from the WAV file.

*B. Time Division*

To deal correctly with our problem, we chose not to encode the timing information in the chromosome although it is possible to implement such approach.

To get the duration of individuals, the target audio signal is analyzed and split using notes onset detection algorithms. The resulting information is a set of time segments. The length of these particular segments represents the duration of each sound that will be synthesized at each corresponding time from the audio signal.

*C. Fast Fourier Transform-FFT*

Applying a straight comparison between two acoustic signals by measuring the difference between the samples is likely to result in similar sounds being rejected because of minute differences in small factors like phase [6]. It would be more convenient to work in the frequency domain. Therefore, phase will be thrown away by the power spectrum, thus enabling differences in the phase of the sine wave making up the signal to be not considered. During the submission of a particular segment of the sound to the ES process, the segment is split in smaller time slots and for each of these slots; a Fast Fourier Transform is applied to compare the magnitudes of all the frequencies with the corresponding slots in the original signal. This comparison happens during the evaluation process, to measure the level of adaptability of an individual.

*D. Fitness Function*

Our fitness function implements the process of comparing each individual with the original signal that needs to be transcribed. In order to do it, at the top level the original signal is first split according to the duration of each composing sound.

We address in this work equal length polyphonic sounds. Those segments are then submitted individually to ES algorithms instances. At the initialisation step, since the audio signal has been split in small segments representing a sound played in a certain interval of time, with our approach the ES will not need to use large size chromosomes, but rather will be constituted of a real-valued vector of only 3 elements or more depending of the number of notes composing a particular sound. These latter ones will enable the ES strategy to optimize only the related segment of the audio signal representing a sound played at a certain interval of time. Therefore, improving the accuracy of the results compared to others methods, which need to encode the full audio signal in a chromosome [5], [6]. At the end of the execution, the results from all ES which took as input each audio segment are returned in the correct order along with their durations.

Before calling the fitness function, the audio segment is again split in smaller time slots of approximately 93 ms by using a window size of 4096 with a samples rate of 44100, resulting in a set of smaller time slices (size 4096). Now, a FFT is applied to each of these time slices along with a Han windowing function to ensure that there are no artefacts of unwanted frequencies as per equation 5.

$$w(n) = 0.5\left(1 - \cos\left(\frac{2\pi n}{N-1}\right)\right) \qquad (5)$$

The same process is applied to the original audio signal that we want to transcribe. The fitness function will then compute its value by summing the distance between each frequency in each time slice of the sound as per equation (6). Since we are minimizing the difference from the frequency magnitude of the original signal with the synthesized one, the closer it will get, the higher will be the level of adaptability.

$$Fitness = \begin{cases} \sum_{t=0}^{tmax} \sum_{f=27.5}^{fmax} \frac{O(t,f)}{X(t,f)}, & X(t,f) < O(t,f) \\ \sum_{t=0}^{tmax} \sum_{f=27.5}^{fmax} \frac{X(t,f)}{O(t,f)}, & X(t,f) \geq O(t,f) \end{cases} \qquad (6)$$

The O(t, f) is the magnitude of frequency f at time slot t in the acoustic audio signal, and X(t, f) is the magnitude of the frequency f at time slot t for each of the synthesized individual. The fitness is computed from time slot t=0 to tmax, going through all times from the beginning to the end, and from fmin=27.5 to fmax=22050 which correspond respectively to the lowest frequency and the nyquist frequency of a 44100 HZ sample rate. The ES will adapt its parameters toward optimal values that will result in the correct transcription of the polyphonic sound.

## V. EXPERIMENTS & RESULTS

The experiments were run using Matlab. We wrote the needed codes for this purpose using this latter one.

The identified constraints of our ES was that the results should be in the interval going from 21 to 108 corresponding to the lowest piano note (MIDI numbers) to 108 which is the highest note of the piano. In order to run our experiments, we synthesized a set of WAV files that we used as our target audio signals needing to be transcribed. We made some assumptions to ease our experiments as follow:

- The smallest duration of a note is 0.5 sec.
- The duration of a sound may be either 0.5 sec., 1 sec., 1.5 sec., 2 sec.
- Notes played at the same time should have the same length.

TABLE II. ES PARAMETERS VALUES

| Parameters | Values |
|---|---|
| Population size | 100 |
| Offspring | 80 |
| Initialization strategy | [0.005, 0.05] |
| Max number of generations | 300 |
| Search space | [21, 108] |

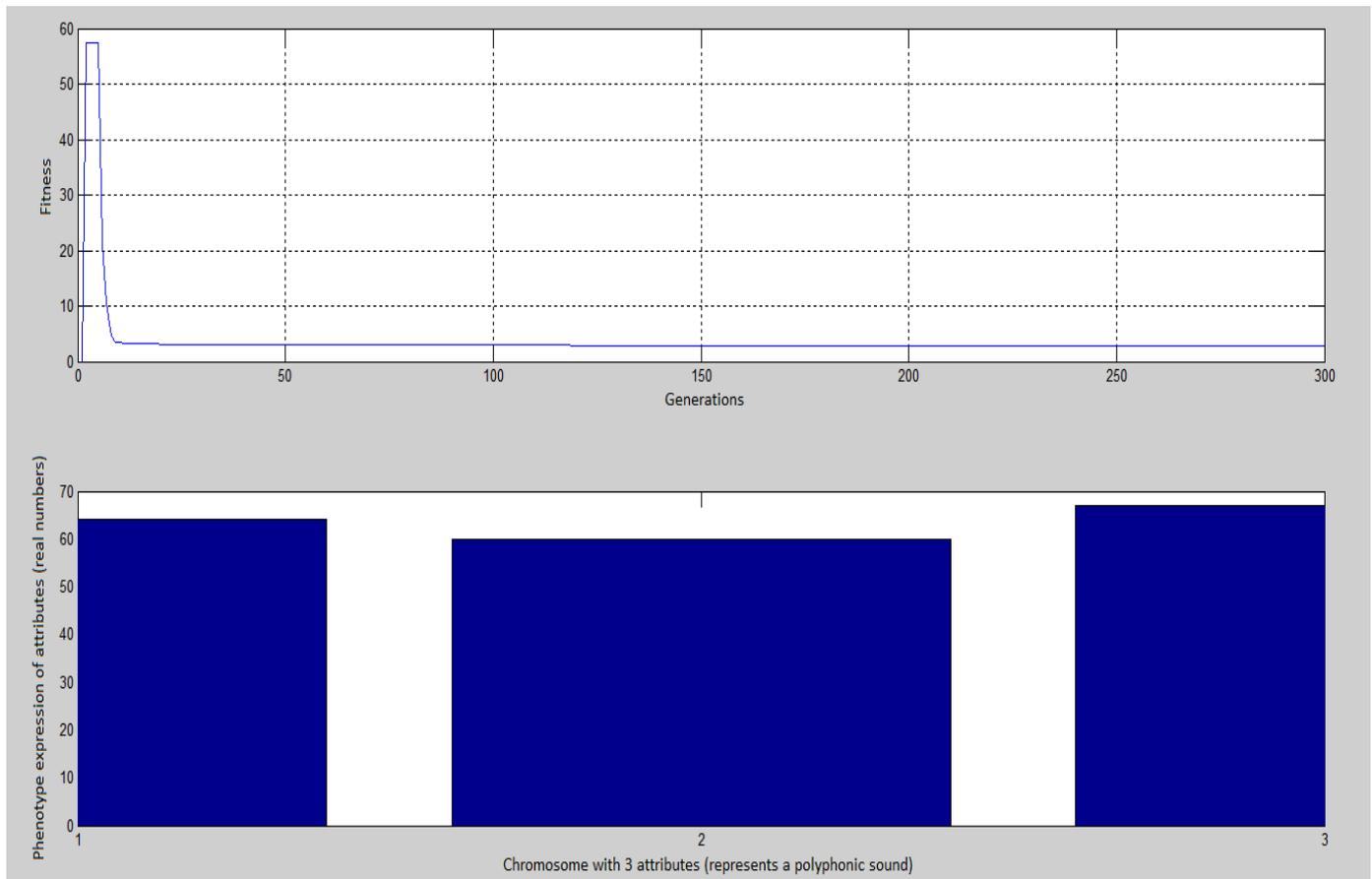

Fig. 1. Graphical Results of transcription of C major chord

Using ES with our approach, the first experiment run was to transcribe a C Major Chord synthesized with the notes C (60)", "G (67)", "E (64)". The transcription was perfectly done as Fig. 1 shows; we got the best results after 9 generations in about 1.5 minute. Figure 1 and Table III presents the best results returned by the ES although it was run for 300 generations without any further improvements of the results

As EA do not guarantee to return the same results even when run different times with the same parameters, it is important to run them many times in order to compare the results as they can produce better ones. Table IV shows the best results obtained after running a second set of tests for the transcription of the same chord (C Major). The results were obtained after 8 generations in approximately 1.5 min.

TABLE III. TRANSCRIPTION OF C MAJOR : SECOND RESULTS

|  | Note 1 | Note 2 | Note 3 |
|---|---|---|---|
| **Rounded numbers** | 64 | 60 | 60 |
| **Real values results** | 63.9985 | 60.1580 | 60.2356 |

We ran some other tests to transcribe a progression of 5 chords as shown in Table V. The ES took only 14 minutes to successfully transcribe them with good accuracy. Compared to others methods using evolutionary algorithms, these results are very interesting. In term of applying machine learning techniques to music transcription, our work clearly shows that ES are also good tools to deal with such problems.

In their work [6], Gustavo Miguel et al reported that the transcription of a progression of 5 chords took about 64 minutes to complete using GAs whereas using ES this shortcoming in term of computation time may be significantly reduced as our results shows.

By increasing the value of the initialization strategy parameter with our approach, a fast convergence with good accuracy can be triggered. Moreover, in our approach we do not submit the full audio signal to the ES which renders the computation very expensive, however we first identify using note onsets detection techniques the duration of each sound and then split the audio signal accordingly. Individual pieces of the audio signal are then submitted independently to ES instances which bring parallelization features to our approach, thus reducing the time for computation.

TABLE IV. CHORDS PROGRESSION OF A POLYPHONIC SOUND

| Chord 1 (1 sec.) | Chord 2 (0.5 sec.) | Chord 3 (0.5 sec.) | Chord 4 (1.5 sec.) | Chord 5 (1 sec.) |
|---|---|---|---|---|
| 60 | 55 | 69 | 64 | 60 |
| 64 | 59 | 76 | 67 | 64 |
| 67 | 62 | 72 | 71 | 67 |

Table VI presents the best results we obtained for the transcription of a 5 chords progression as returned by the ES without applying rounding process. The results are presented as real numbers. And as we can observe, the transcription process returns the results with a good accuracy. The experiments we have done on the application of ES in polyphonic music transcription show that these machine learning tools are good tools that can be applied in polyphonic music transcription. A tradeoff therefore on the value of the strategy parameters exists as in certain conditions, this latter one should be increased to obtained good results while on the other side some problems will need to decrease this value to obtain good results in term of optimization. Different values of this parameter should be tested and tuned to an acceptable one.

TABLE V. RESULTS OF TRANSCRIPTION PROGRESSION

| Chord 1 (1 sec.) | Chord 2 (0.5 sec.) | Chord 3 (0.5 sec.) | Chord 4 (1.5 sec.) | Chord 5 (1 sec.) |
|---|---|---|---|---|
| 60.0437 | 55.2045 | 68.8471 | 64.0000 | 60.3624 |
| 63.9456 | 59.1467 | 76.0000 | 66.9162 | 64.0132 |
| 66.8463 | 62.1978 | 72.4153 | 70.1457 | 66.2682 |

## VI. CONCLUSION

We presented in this paper a novel approach for automatic polyphonic music transcription using evolution strategies. In this context, we showed through our experiments that the frequencies lattice of a perceived sound in an audio signal can be reconstructed by optimization procedures as the transcription of polyphonic sound can also be addressed as a search space problem. The application of ES as learning tools in automatic polyphonic music transcription has proven itself to be adapted for such problem in comparison to other state-of-the-art techniques. Moreover, using these machine learning tools, some shortcomings presented by other evolutionary algorithms approaches that include the long computation time and the high computation cost can be solved. As evolution strategies use self-adapting parameters for optimization, we showed that tuning the strategy parameter that controls the standard deviation, may trigger a fast convergence toward the optima while reducing the computation time. At the other side, the computation cost and time is addressed by mean of parallelization approach to ensure that the workload is distributed among the available resources, thus reducing the computation time.

An interesting and potential direction will be to investigate on the behavior of those machine learning tools when it comes to acoustic sounds. A question can be raised at this level as MIDI numbers are integer values that represent the pitches. However, knowing that it is hard for an acoustic instrument which is manually tuned by a human to sound exactly as the correspondent target frequency, but only, it can get near the ideal frequency. Investigating further on how ES can be improved to transcribe sounds from acoustic instruments constitute a potential challenge as its solution can only be seen in a search space that should include frequencies representations with their real values.

REFERENCES

[1] Herve Kabamba Mbikayi, "An Evolution Strategy Approach toward Rule-set Generation for Network Intrusion Detection Systems (IDS)," International Journal of Soft Computing and Engineering (IJSCE), vol. 2 Issue-5, pp. 201–205, 2012.

[2] Trung Hau Tran, Cédric Sanza, Yves Duthen, "Evolving Prediction Weights Using Evolution Strategy," Proceedings of GECCO conference companion on Genetic and evolutionary computation, pp. 2009-2016, 2008.

[3] Alain de Cheveigné, Hideki Kawahara, "Multiple period estimation and pitch perception model," Speech Communication 27, 175-185.

[4] Manuel Davy and Simon J. Godsill, "Bayesian Harmonic Models for Musical Signal Analysis Bayesian Statistics VII," Oxford University Press, 2003.

[5] G Reis, F Fernandez, A Ferreira, "Genetic algorithm approach to polyphonic music transcription for MIREX 2008," in Proc of the 4th Music Information Retrieval Evaluation eXchange (MIREX), 2008.

[6] Gustavo Miguel Jorge dos Reis, Francisco Fernandez de Vega, " Electronic Synthesis using Genetic Algorithms for Automatic Music Transcription," GECCO'07, July 7–11, 2007.

[7] Masataka Goto, "A Real-time Music-scene-description System: Predominant-F0 Estimation for Detecting Melody and Bass Lines in Real-world Audio Signals," Speech Communication (ISCA Journal), Vol.43, No.4, pp.311-329, September 2004.

[8] K. Kashino, K. Nakadai, T. Kinoshita, and H. Tanaka, "Organization of hierarchical perceptual sounds: Music scene analysis with autonomous processing modules and a quantitative information integration mechanism," In IJCAI, pp. 158-164, 1995.

[9] Rechenberg, I., "Evolutionsstrategie: Optimierung technischer Systeme und Prinzipien der biologische," Frommann-Holzboog, Stuttgart, 1973.

[10] Schwefel, H.-P. Numerical Optimization of Computer Models, Chichester: Wiley, 1981.

[11] A. P. Klapuri, "Automatic music transcription as we know it today," Journal of New Music Research, vol. 33(3), pp. 269-282, 2004.

[12] G. Garcia, "A genetic search technique for polyphonic pitch detection," In Proceedings of the International Computer Music Conference (ICMC), Havana, Cuba, 2001

[13] D. Lu, "music transcription using genetic algorithms and electronic synthesis," Computer Science Undergraduate Research, University of Rochester, New York, USA

[14] Gustavo Reis, Francisco Fernandéz, Annibal Ferreira, "A genetic Algorithm for polyphonic Transcription of Music," In proceeding of: Proceedings of the 2007 Doctoral Symposium on Artificial Intelligence, 2007

[15] Emmanouil Benetos, Simon Dixon, Dimitrios Giannoulis, Holger Kirchhoff, and Anssi Klapuri, 13[th] International Society for Music Information Retrieval Conference, 2012.

[16] H. Kirchhoff, S. Dixon, and A. Klapuri, Shift-variant nonnegative matrix deconvolution for music transcription, In ICASSP, 2012.